\documentclass[sigconf]{acmart}
\definecolor{blue}{HTML}{1F77B4}
\definecolor{orange}{HTML}{FF7F0E}
\definecolor{green}{HTML}{2CA02C}

\AtBeginDocument{%
  \providecommand\BibTeX{{%
    \normalfont B\kern-0.5em{\scshape i\kern-0.25em b}\kern-0.8em\TeX}}}

\usepackage{geometry}
\usepackage{amsmath}
\usepackage{amsfonts}
\usepackage{booktabs} % For pretty tables
\usepackage{caption} % For caption spacing
\usepackage{subcaption} % For sub-figures
\usepackage{graphicx}
\usepackage{algorithm}
\usepackage{algorithmic}
\usepackage{pgfplots}
\usepackage{soul}
\usepackage[all]{nowidow}
\usepackage[utf8]{inputenc}
\usepackage{tikz}
\usetikzlibrary{er,positioning,bayesnet}
\usepackage{multicol}
\usepackage{hyperref}
\usepackage[inline]{enumitem} % Horizontal lists
\copyrightyear{2022}
\acmYear{2022}
\setcopyright{acmlicensed}\acmConference[BCB '22]{13th ACM International Conference on Bioinformatics, Computational Biology and Health Informatics}{August 7--10, 2022}{Northbrook, IL, USA}
\acmBooktitle{13th ACM International Conference on Bioinformatics, Computational Biology and Health Informatics (BCB '22), August 7--10, 2022, Northbrook, IL, USA}
\acmPrice{15.00}
\acmDOI{10.1145/3535508.3545539}
\acmISBN{978-1-4503-9386-7/22/08}

\begin{document}
\title{Distribution-based Sketching of Single-Cell Samples}
\author{Vishal Athreya Baskaran}
\email{athreya@cs.unc.edu}
\affiliation{%
  \institution{Department of Computer Science, UNC-Chapel Hill
}
\country{USA}
}

\author{Jolene Ranek}
\email{ranekj@live.unc.edu}
\affiliation{%
  \institution{Computational Medicine Program, Curriculum in Bioinformatics and Computational Biology, UNC-Chapel Hill
}
\country{USA}
}

\author{Siyuan Shan}
\email{siyuanshan@cs.unc.edu}
\affiliation{%
  \institution{Department of Computer Science, UNC-Chapel Hill
}
\country{USA}
}

\author{Natalie Stanley}
\email{natalies@cs.unc.edu}
\affiliation{%
  \institution{Computational Medicine Program, Department of Computer Science, UNC-Chapel Hill
}
\country{USA}
}

\author{Junier B. Oliva}
\email{joliva@cs.unc.edu}
\affiliation{%
  \institution{Department of Computer Science, UNC-Chapel Hill
}
\country{USA}
}

\renewcommand{\shortauthors}{Baskaran et al.}
\begin{abstract}
Modern high-throughput single-cell immune profiling technologies, such as flow and mass cytometry and single-cell RNA sequencing can readily measure the expression of a large number of protein or gene features across the millions of cells in a multi-patient cohort. While bioinformatics approaches can be used to link immune cell heterogeneity to external variables of interest, such as, clinical outcome or experimental label, they often struggle to accommodate such a large number of profiled cells. To ease this computational burden, a limited number of cells are typically \emph{sketched} or subsampled from each patient. However, existing sketching approaches fail to adequately subsample rare cells from rare cell-populations, or fail to preserve the true frequencies of particular immune cell-types. Here, we propose a novel sketching approach based on Kernel Herding that selects a limited subsample of all cells while preserving the underlying frequencies of immune cell-types. We tested our approach on three flow and mass cytometry datasets and on one single-cell RNA sequencing dataset and demonstrate that the sketched cells (1) more accurately represent the overall cellular landscape and (2) facilitate increased performance in downstream analysis tasks, such as classifying patients according to their clinical outcome. An implementation of sketching with Kernel Herding is publicly available at \url{https://github.com/vishalathreya/Set-Summarization}.
\end{abstract}


\maketitle              % typeset the header of the contribution
\section{Introduction}

Advances in high throughput single-cell technologies have transformed our understanding of cellular heterogeneity across a range of biological and clinical applications. For example, single-cell measurements have been used to construct whole organism cell atlases \cite{HumanCellAtlas,devAtlas} that can serve as prototypical healthy references that assist in the discovery of novel disease cell states \cite{scArches}. Translationally, single-cell modalities have provided insight into complex immune responses linked to clinical outcomes, such as surgical recovery \cite{ganio}, pregnancy \cite{immuneClock}, tumor heterogeneity \cite{phenograph}, and infectious disease \cite{covidCytof}. 

\begin{figure*}[t!]
\begin{center}
  \includegraphics[width=1.3\columnwidth]{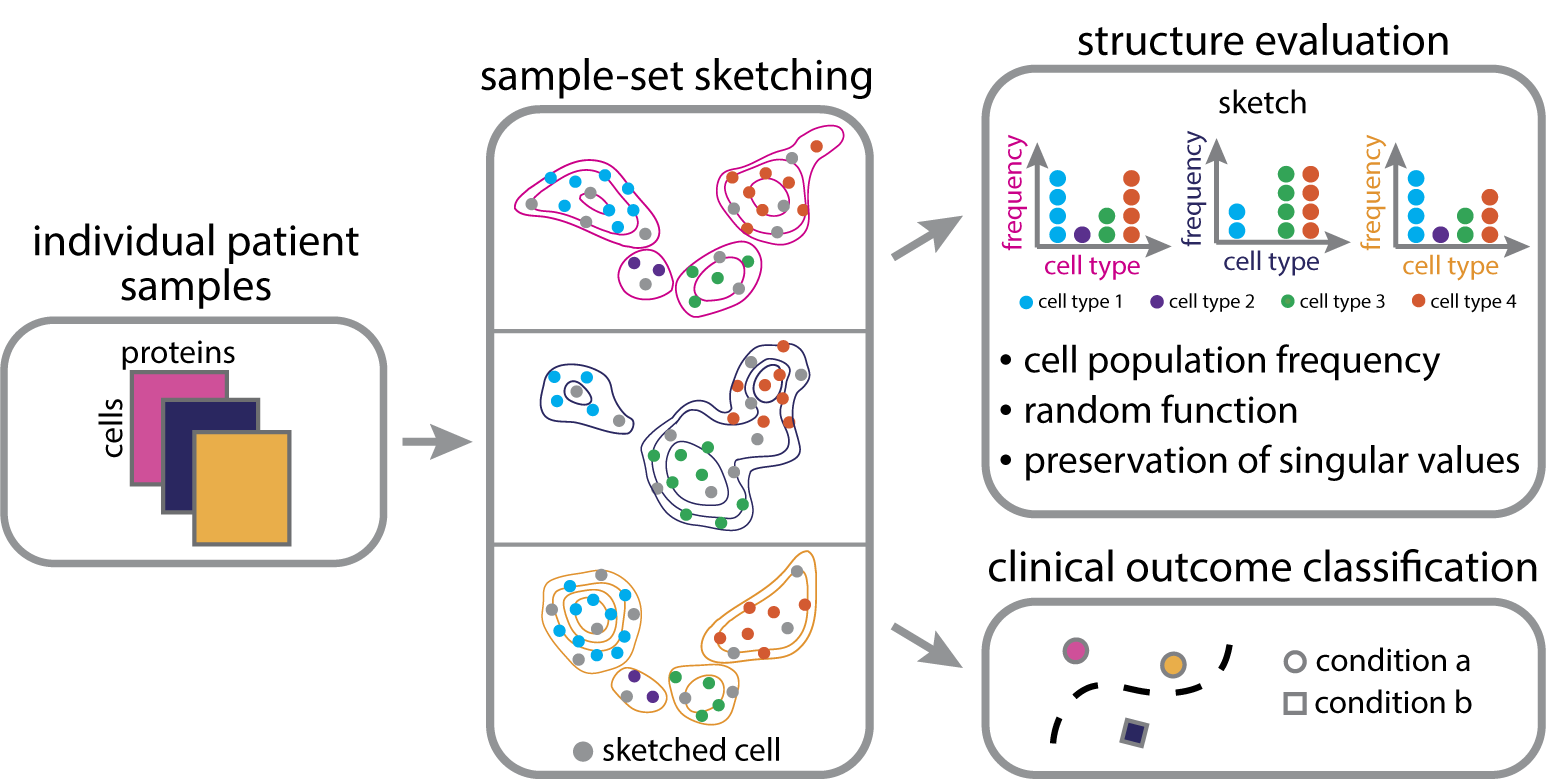}
\caption{{\bf Overview of Kernel Herding for Sketching Single-Cell Data.} Multiple \emph{sample-sets} are created after profiling the expression of several genomic or protein features in single cells across multiple individuals. We used Kernel Herding to select a limited representative subset of cells from sample-sets for each individual. Our approach was then evaluated in terms of the quality of cellular landscape preservation and classification performance of sample-sets according to the clinical outcomes of the associated patients.}
\label{fig:fig1}
\end{center}  
\end{figure*}

Single-cell flow and mass cytometry (e.g. CyTOF) technologies have been particularly useful for gaining an in-depth understanding of the diversity of immune cell-types and their relation to disease, as they simultaneously profile 20-45 proteins collectively across all of the cells profiled in a multi-patient cohort \cite{bendall1,spitzer,perplexed}. This data can be leveraged in statistical models for identifying differentially abundant cell-populations \cite{diffcyt,cydar,milo}, kernel density estimation methods to characterize treatment resistant cells \cite{meld}, and classification algorithms to predict a patient’s clinical outcome \cite{cytoset}. Similarly, genomic technologies such as single-cell RNA sequencing (scRNA-seq) technologies can profile $20,000$ or more gene expression measurements in thousands of cells \cite{scRNA}.  Given the large number of cells measured in such datasets, downstream analysis tasks present a computational challenge and quickly become intractable. Traditional approaches for reducing the size of input data, such as uniform downsampling or clustering, may fail to identify rare but clinically meaningful cell-populations. This limits both the utility of these algorithms and the ability to obtain meaningful results. 

Two alternative sketching-based approaches have been developed to intelligently select a subset of cells such that the overall cellular landscape of the full data is preserved. Geometric Sketching first approximates the underlying geometry of the data through a covering of equal volume hypercubes, then evenly samples points from each at random \cite{berger2}. Alternatively, Hopper constructs a sketch by iteratively minimizing the Hausdorff distance to ensure points in the full data are well-represented in the sketch \cite{berger1}. 
Both methods provide an even sampling of cells that focuses on capturing rare cell-types.
However, as a result, they fail to preserve the overall distributions of cell-type frequencies which can be crucial for clinical outcome prediction and biological interpretation \cite{ganio,Mathew2020-kb}.
In a philosophical divergence with this previous sketching work, here we prioritize the sketch's ability to act as a \emph{stand-in} for the original sample-set. That is, here we consider sketches that obtain \emph{similar downstream outcomes} to the original sample-sets when analyzed. 
In some respects, this goal is a generalization of prior formulations, since rare-cell types must be preserved if the sketch can act as a stand-in. However, our formulation also allows for the preservation of general downstream analysis, which may not be entirely dependent on rare-cell types. 
Thus, this work presents a novel, more direct approach to sketching single-cell samples.
%Although both methods provide an even sampling of cells to better capture rare cell-types, they fail to preserve the overall distributions of cell-type frequencies which can be crucial for clinical outcome prediction and biological interpretation.
% Therefore, there is a critical need to appropriately sketch cells, such that 1) cells from rare populations are adequately represented and 2) the frequencies of cells across immune cell-types are preserved. 

We propose to preserve downstream analysis of a single-cell sample by explicitly preserving the distribution in the sketched sample. In this work, we show that this can be achieved via Kernel Herding \cite{chen2010super}, a kernel-based approach to sub-sampling. Furthermore, we show how to efficiently compute Kernel Herding sketches through random Fourier features \cite{rahimi2007random}, which allows our sketch approach to scale to large sample-sets. We first evaluate our approach on three diverse flow and mass cytometry datasets by computing several statistics to quantify cellular landscape preservation and performance in downstream tasks, such as patient classification according to clinical outcome. Furthermore, we demonstrate how this approach can be used across single-cell data modalities with a single-cell RNA sequencing application. Our Kernel Herding sketching approach (outlined in Fig. \ref{fig:fig1}) now makes it possible to select a limited subset of cells from each profiled sample-set that adequately represents immune cell-types and their relative frequencies. 

\section{Methods}
\label{sec:methods}
%\textcolor{red}{Junier and Vishal} \\ % JO: I was hoping Vishal would have provided more text here

% JO: I have no idea what the notation that is to be used later is so I'll just put down w/e and it can be adjusted afterward

Here we are considering the problem of finding a representative subset to an original sample-set of cells $\mathcal{X} = \{x_i \in \mathbb{R}^d \}_{i=1}^n$. 
If subsets are to act as a stand-in for original sample-sets, then they should provide similar outcomes when processed as the original sets.
Thus, we consider a subset, $\hat{\mathcal{X}} = \{\hat{x}_i\}_{i=1}^m \subset \mathcal{X}$, to be representative if it maintains the distribution of the original set $\mathcal{X}$ and yields similar results to $\mathcal{X}$  when processed. \\

% \noindent {\bf Notation for sample-sets, etc} \\

% See NSF proposal, but look at the sets of cells across each patient and notation for the sets we are ultimately trying to construct \\

\noindent {\bf Problem Formulation}
% We wish to take our large sample-sets and reduce them to a limited subset of cells that are representative of the overall cellular landscape.... blah blah blah \\
At an abstract level, we shall construct a subset $\hat{\mathcal{X}}$ whose statistics match the original sample-set $\mathcal{X}$. Of course, for any finite number of statistics, there may be an ambiguity over the original distribution. For instance, a large variety of samples may share the same mean, thus simply finding a subset of a similar mean as the original set will not suffice to ensure a representative subsample. Instead, we shall use the kernel mean embedding \cite{muandet2017kernel} as the target `statistic' to match in the subset. We briefly explain kernels and kernel mean embedding below. \\

\noindent {\bf Kernel Mean Embedding}
Kernel methods have been successfully applied in a myriad of machine learning problems such as  regression \cite{vovk2013kernel}, classification \cite{cortes1995support}, and dimensionality reduction \cite{mika1998kernel}.
Underlying kernel methods is a positive definitive kernel function $K : \mathbb{R} \times \mathbb{R} \mapsto \mathbb{R}$\,\footnote{Kernels may also generalize over non-real domains}, which induces a reproducing kernel Hilbert space (RKHS) (e.g.~see \cite{berlinet2011reproducing} for further details). 
In addition to working over typical vector-values data, recently kernels have also been deployed to operate over distributions (e.g.~\cite{muandet2017kernel}). The \emph{kernel mean embedding} $\mu_p: \mathbb{R} \mapsto \mathbb{R}$ represents a distribution, $p$, as:
\begin{equation}
    \mu_p(\cdot) = \mathbb{E}_{x\sim p}[k(x, \cdot)]
\end{equation}
Note that $\mu_p$ is itself a function, defined by the expected kernel evaulation to points drawn from a distribution $p$. For a special class of `\emph{characteristic}' kernels, $k$, such as the common radial-basis function (RBF) kernel $k(x, x^\prime) = \exp(-\frac{1}{2 \gamma} ||x - x^\prime||^2)$, $||\mu_p - \mu_q|| = 0$ if and only if $p = q$. Thus, for characteristic kernels the kernel mean embedding will be unique to its distribution and matching the kernel embedding exactly guarantees that one matches a distribution. In general, the distance\footnote{Corresponding to the RKHS norm} $||\mu_p - \mu_q||$ induces a divergence, the maximum mean discrepancy (MMD) \cite{gretton2008kernel}, between distributions, which has been useful in comparing distributions (e.g.~\cite{gretton2012kernel}).

In this work we sketch a sample-set with a subset $\hat{\mathcal{X}} \subset \mathcal{X}$ that preserves the empirical distribution found in the original sample-set $\mathcal{X}$ with the mean embedding $\mu_{\mathcal{X}}$ where:
\begin{equation}
    \mu_{\mathcal{X}}(\cdot) = \frac{1}{n} \sum_{i=1}^n k(x_i, \cdot).
\end{equation}
That is, we look for a subset $\hat{\mathcal{X}}$ whose empirical distribution has a small MMD from that of the original set $\mathcal{X}$, $||\mu_{\hat{\mathcal{X}}} - \mu_\mathcal{X}||$.
Note that a uniform subsample $\mathcal{U} \subset \mathcal{X}$ provides a reasonable approximation when the cardinality $|\mathcal{U}|$ is large enough. That is $\mu_{\mathcal{U}}(\cdot) = \frac{1}{|\mathcal{U}|} \sum_{x \in \mathcal{U}} k(x, \cdot)$, shall approximate $\mu_{\mathcal{X}}$ when using enough uniformly subsampled points. However, as shown by \cite{chen2010super}, one may get a more representative subsample through \emph{Kernel Herding} as described below. \\

\noindent 
\noindent{\bf Kernel Herding }
Kernel herding is a greedy algorithm to approximate a mean embedding $\mu_p$ with a subset of a set of samples $\mathcal{X} \sim p$, using a subset of points $\hat{\mathcal{X}} = \{\hat{x}_i\}_{i=1}^m \subset \mathcal{X}$. Here, we consider the empirical distribution $p = \frac{1}{n} \sum_{i=1}^n \delta(x_i)$, where $\delta$ is the Dirac-delta function. Hence, we look to approximate $\mu_\mathcal{X}$ with $\mathcal{X}$ being our original sample.
Kernel herding is especially applicable for synthesizing a large number of cells, due to the following properties. First, the rate of convergence for the error $\| \mu_\mathcal{X} - \frac{1}{m} \sum_{j=1}^m k(\hat{x}_i, \cdot) \|^2$ is $\mathcal{O}(\frac{1}{m})$, which is much faster than the $\mathcal{O}(\frac{1}{\sqrt{m}})$ rate for a uniformly random subsample. That is, {Kernel Herding can provide a synthesized subset of $\sqrt{m}$ points that approximates the target distribution as well as $m$ uniformly subsampled points}. 
\ul{This property implies that the Kernel Herding subset shall maintain the distributional properties found in the original sample-set.} 
Second, it can be shown that the Kernel Herding subset $\hat{\mathcal X}$ also approximates the expectation of any function $h$\footnote{In the corresponding reproducing kernel Hilbert space (RKHS).}, $\mathbb{E}_{x \sim p}[h(x)]$, with a mean over $\hat{\mathcal{X}}$, $\frac{1}{m} \sum_{j=1}^m h({\hat x}_j)$. \ul{This property enables one to obtain similar downstream models when using the summarized subset as the expectation of functions evaluated}.

The Kernel Herding subset $\hat{\mathcal{X}}$ is computed as follows:

  \begin{algorithm}[H]
        \caption{\textsc{Compute Subsampled Set using Kernel Herding}}
        \begin{algorithmic}[1]
        \label{algo_kh}
          \REQUIRE A set of cells $\mathcal{X}$, from which the number of cells subsampled is $m$, Radial-Basis Function kernel $k$.

\STATE Initialize $j \leftarrow 1$, $\mathcal{\hat{X}} \leftarrow \emptyset$

\WHILE{$j \leq m$}
    \STATE $\hat{\mathbf{x}} \leftarrow \underset{x \in \mathcal{X}}{\mathrm{argmax}} \ \dfrac{1}{N} \sum_{x^\prime \in \mathcal{X}} k(x, x^\prime) - 
    \dfrac{1}{j} \sum_{x^\prime \in \hat{\mathcal{X}}} k(x, x^\prime)$
    \STATE $\mathcal{\hat{X}} \leftarrow \mathcal{\hat{X}} \cup \{\hat{\mathbf{x}}\}$
    \STATE $j \leftarrow j + 1$
\ENDWHILE
\STATE \textbf{return} $\hat{\mathcal{X}}$

\end{algorithmic}
\end{algorithm}

\noindent 
\noindent {\bf Random Fourier Features }
Performing Kernel Herding with a kernel function $k$, however, may not scale since an iteration requires an $O(n^2)$ computation (equivalent to computing the Gram matrix of pairwise kernel evaluations on $\mathcal{X}$). To rectify this, we propose to use random Fourier frequency features \cite{rahimi2007random}. For a shift-invariant kernel (such as the RBF kernel), random Fourier features provide a feature map $\varphi(x) \in \mathbb{R}^D$ such that the dot product in feature space approximates the kernel evaluation, $\varphi(x)^T \varphi(x^\prime) \approx k(x, x^\prime)$ (e.g.~see \cite{rahimi2007random,oliva2014fast,rahimi2007random} for further details). Using the dot product of $\varphi(x)$, our mean embedding becomes $\mu_\mathcal{X} = \frac{1}{n} \sum_{i=1}\varphi(x_i)$, where $\mu_\mathcal{X}(x^\prime) = \frac{1}{n} \sum_{i=1}\varphi(x_i)^T \varphi(x^\prime)$.
In practice, the random features are constructed by drawing frequencies, $W \in \mathbb{R}^{d \times \frac{D}{2}}$, at random (\textit{iid} column-wise) once, and holding them fixed to construct the features 
\begin{equation}
    \label{eq:randfeat}
    \varphi_W(x) = [\sin(\mathbf{W}^T x), \cos(\mathbf{W}^T x)] \in \mathbb{R}^D,
\end{equation}
where $[\cdot,\cdot]$ denotes concatenation.
For instance, to approximate the RBF kernel, $k(x, x^\prime) = \exp(-\frac{1}{2 \gamma} ||x - x^\prime||^2)$, we would draw the $d$-dimensional frequencies $W_i \overset{iid}{\sim} \mathcal{N}(0, \gamma^{-1} I)$ \cite{rahimi2007random}.
It is simple to show that when using the approximation $\varphi_W(x)^T \varphi_W(x^\prime) \approx k(x, x^\prime)$, one may compute Algorithm \ref{algo_kh} with $O(n)$ iterates by avoiding pairwise kernel computations \cite{chen2010super}; we detail this below.
\noindent 
  \begin{algorithm}[H]
        \caption{\textsc{Compute Subsampled Set using RBF Kernel Herding with Random Features}}
        \begin{algorithmic}[1]
        \label{algo_kh_rff}
          \REQUIRE A set of cells $\mathcal{X}$ with dimensionality $d$, from which the number of cells subsampled is $m$, dimensionality of the random feature space $D$ and kernel hyperparameter $\gamma$.
          
\STATE \texttt{\# Draw random Fourier frequencies}
\STATE Compute $\mathbf{W}\in \mathbb{R}^{d \times \frac{D}{2}}$ by sampling its elements independently $\mathbf{W}_{i,j} \sim \mathcal{N}(0, \frac{1}{\gamma})$ 

\STATE  \texttt{\# Subsampling using Kernel Herding}

\STATE Initialize $j \leftarrow 1$, $\mathcal{\hat{X}} \leftarrow \emptyset$, $\theta_0 \leftarrow \frac{1}{n}\sum_{i=1}^{n}\varphi_{\scriptscriptstyle \mathbf{W}}(x_i)$

\WHILE{$j \leq m$}
    \STATE $\hat{\mathbf{x}} \leftarrow \underset{x \in \mathcal{X}}{\mathrm{argmax}} \ \theta_{j-1}^T \varphi_{\scriptscriptstyle \mathbf{W}}(x_i)$
    \STATE $\mathcal{\hat{X}} \leftarrow \mathcal{\hat{X}} \cup \{\hat{\mathbf{x}}\}$
    \STATE $\theta_j \leftarrow \theta_{j-1} + \theta_0 - \varphi_{\scriptscriptstyle \mathbf{W}}(\hat{\mathbf{x}})$
    \STATE $j \leftarrow j + 1$
\ENDWHILE
\STATE \textbf{return} $\hat{\mathcal{X}}$

\end{algorithmic}
\end{algorithm}

\subsection{Single-Cell Datasets Used in Experiments}
In all experiments, we used publicly available, multi-sample flow, mass cytometry (CyTOF), and single-cell RNA sequencing datasets. For the flow and mass cytometry datasets, each sample-set consists of a collection of protein markers (up to $\sim$45) measured across individual cells. For the single-cell RNA sequencing dataset, each sample-set consists of a collection of gene expression measurements ($\sim$20k) measured across individual cells. Here, we briefly introduce the multi-sample publicly available datasets used in our experiments. All preprocessed data are available in the Zenodo repository: \url{https://zenodo.org/record/6546964}.

\begin{figure*}[t!]
    \centering
  \includegraphics[width=2\columnwidth]{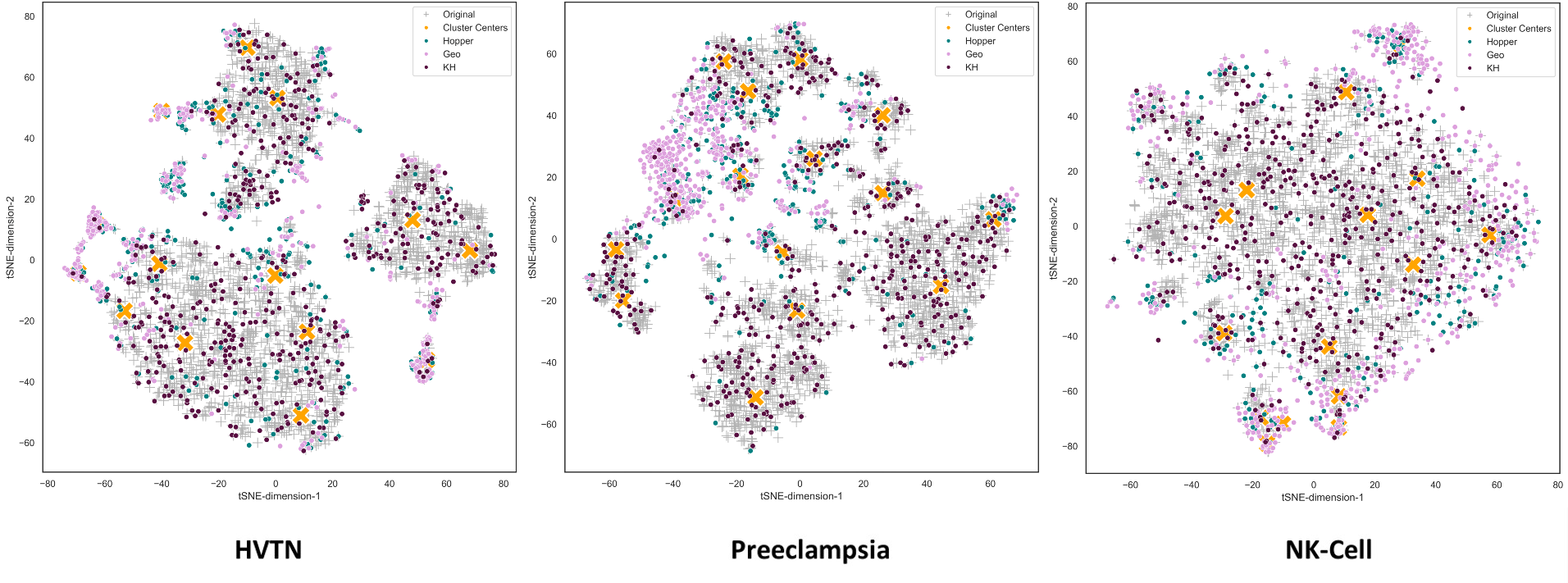}
\caption{\bf t-SNE visualizations of the sketches produced under each method. The distribution of points sketched with Kernel Herding (KH) (dark purple points) follows that of the original data (gray, background points representing a large number of randomly sampled cells). Sketches with Kernel Herding adequately represent rare populations and retain their relative frequencies, while Geometric Sketching and Hopper struggle with these aspects.}
    \label{fig:cluster_tsne}
\end{figure*}

\begin{itemize}[leftmargin=10pt]
\item  {\bf Preeclampsia}
The preeclampsia CyTOF dataset \cite{pree} includes samples collected longitudinally from 12 healthy women and 11 women with preeclampsia throughout their pregnancies. All patient samples were downloaded from Flow Repository under Repository ID FR-FCM-ZYRQ (\url{http://flowrepository.org/id/FR-FCM-ZYRQ}). Classification tasks distinguished between healthy from preeclamptic women.

\item {\bf HVTN}
The HIV Vaccine Trials Network (HVTN) is a Flow Cytometry dataset and consists of 96 total sample-sets of T-cells that were each subjected to stimulation with either Gag or Env \cite{flowcap} (downloaded from Flow Repository, ID FR-FCM-ZZZV \url{http://flowrepository.org/id/FR-FCM-ZZZV}). Classification tasks distinguished Gag from Env stimulated samples

\item {\bf NK-Cell}
The NK-Cell CyTOF dataset profiled NK-Cells across 21 individuals who were either positive or negative for Cytomegalovirus (CMV) \cite{cellCNN} (downloaded from \url{https://github.com/eiriniar/CellCnn}). Classification tasks distinguished CMV positive (CMV+) from CMV negative (CMV-) samples.

\item {\bf MS}
The multiple sclerosis (MS) single-cell RNA sequencing dataset \cite{Schafflick} consists of peripheral blood samples collected from 4 MS patients and 4 healthy controls. Patient samples were accessed from the Gene Expression Omnibus using the accession code GSE138266. We performed standard single-cell RNA sequencing data preprocessing, including filtering cells according to read depth and distribution of molecular counts, removing cells with greater than 20 percent mitochondrial transcripts, and retaining genes that were expressed amongst a minimum of 5 cells. Following quality control filtering, we normalized the data to account for differences in sequencing depth by estimating size factors using Scran pooling normalization v1.20.1 \cite{scran} and scaling them across batches using Batchelor v1.8.0. We then performed batch effect correction using ComBat \cite{combat}. Lastly, we restricted the feature space by selecting for highly variable genes on \path{log + 1} transformed data using a normalized dispersion measure in Scanpy v1.8.1 (flavor = Seurat, minimum mean = 0.012, minimum dispersion = 0.25, maximum mean = 5). In our subsequent experiments, we performed principal component analysis on the preprocessed dataset and used the top 50 components for downstream tasks. 
%We performed PCA on the preprocessed dataset and used 50 components. 
\end{itemize}

\section{Results}

We compared the performance of Kernel Herding to sketches obtained using Geometric Sketching, Hopper, and IID subampling for tasks related to the overall preservation and usefulness of the resulting immunological landscape. That is, we sought to evaluate whether or not the sketches adequately represented all major immune cell-types and their relative frequencies and could be used to produce meaningful immunological features for downstream tasks. Code for reproducing the results of all subsequent experiments is publicly available at \url{https://github.com/vishalathreya/Set-Summarization}.

\subsection{Description of Related Algorithms}

Here, we briefly define the sketching approaches that were compared to in our experiments.
\begin{itemize}[leftmargin=10pt]
\item {\bf Geometric Sketching}
Geometric Sketching introduced in Ref. \cite{berger2} infers a \emph{plaid covering} of cells in the high-dimensional space. Cells are sketched through volume-dependent sampling by selecting the same number of cells from sections of the high-dimensional plaid covering. 

\item {\bf Hopper}
Hopper introduced in Ref. \cite{berger1} forms a sketch by using fastest first traversal, which is a greedy approximation to the $k$-center problem. Intuitively, this sketching approach sequentially adds cells to the sketch that are sufficiently different from those that were already included in the sketch.

\item {\bf Independent and Identically Distributed Subsampling (IID)}
IID sketches were generated by simply selecting a random subsample of cells from each sample-set. Here, each cell had the same probability of being selected in the sketch. 
\end{itemize}

\subsection{t-SNE Visualizations of Sketched Regions of the Cellular Landscape}
We begin with a qualitative assessment of sketching approaches on cytometry data (Fig.~\ref{fig:cluster_tsne}).
In order to get a visual understanding of the regions of the cellular landscape included in the sketches produced by each method, we plotted a $2d$ t-SNE projection of respective sketches (colored points) along with a set of overall cells from a large sub-sample of the original set (gray-colored points). 
I.e. Fig.~\ref{fig:cluster_tsne}, plots cells from three samples, one for each respective dataset (similar results may be obtained from other samples).
Furthermore, we projected cluster centroids (15, from k-means) as large yellow crosses.
Taken together, the respective plots give an overview of the cells found in each original sample-set (gray), and which cells were then included in sketches (colored).
Note that an IID sketch would stem from a uniform sub-sample of the gray points, which was omitted for visual clarity.
One can observe that the Geometric Sketching (\textit{Geo}, pink, Fig.~\ref{fig:cluster_tsne}) sketches concentrate over a few sub-regions of the cellular-space, leaving large regions of cells underrepresented in the sketch. This is also true of Hopper (\textit{Hopper}, green, Fig.~\ref{fig:cluster_tsne}), though to slightly lesser extent. Finally, we can observe that our proposed Kernel Herding (\textit{KH}, purple, Fig.~\ref{fig:cluster_tsne}) sketches yield a more representative coverage of the cell-space. 
Following our philosophical goal of obtaining sketches that may act as general \emph{stand-ins} for original sample-sets, it is intuitive that the discrepancy in representation of sketches to the original shall result in discrepancies of outputs of downstream analysis. Moreover, following the insights gained from Kernel Herding (Sec.~\ref{sec:methods}), it is intuitive that IID sketches shall be less representative than those from Kernel herding. In subsequent experiments below, we show that these intuitions hold empirically.
%We found that sketches produced through Kernel Herding adequately preserve all regions of the cellular landscape and their relative frequencies, including rare regions. 
% In contrast, sketches produced by Geometric Sketching and Hopper tend to include an overabundance of cells in particular regions (e.g. light purple points corresponding to geometric sketching), while achieving very sparse representations of other cell-types. While Hopper may be more robust to this issue and covers the overall cellular landscape better (green points), its corresponding sketches do not seem to adequately preserve cell-frequencies. On the other hand, Geo-Sketch samples form concentrated clusters in few regions while leaving other regions untouched.

\begin{figure}[H]
    \centering
    \includegraphics[width=.85\columnwidth]{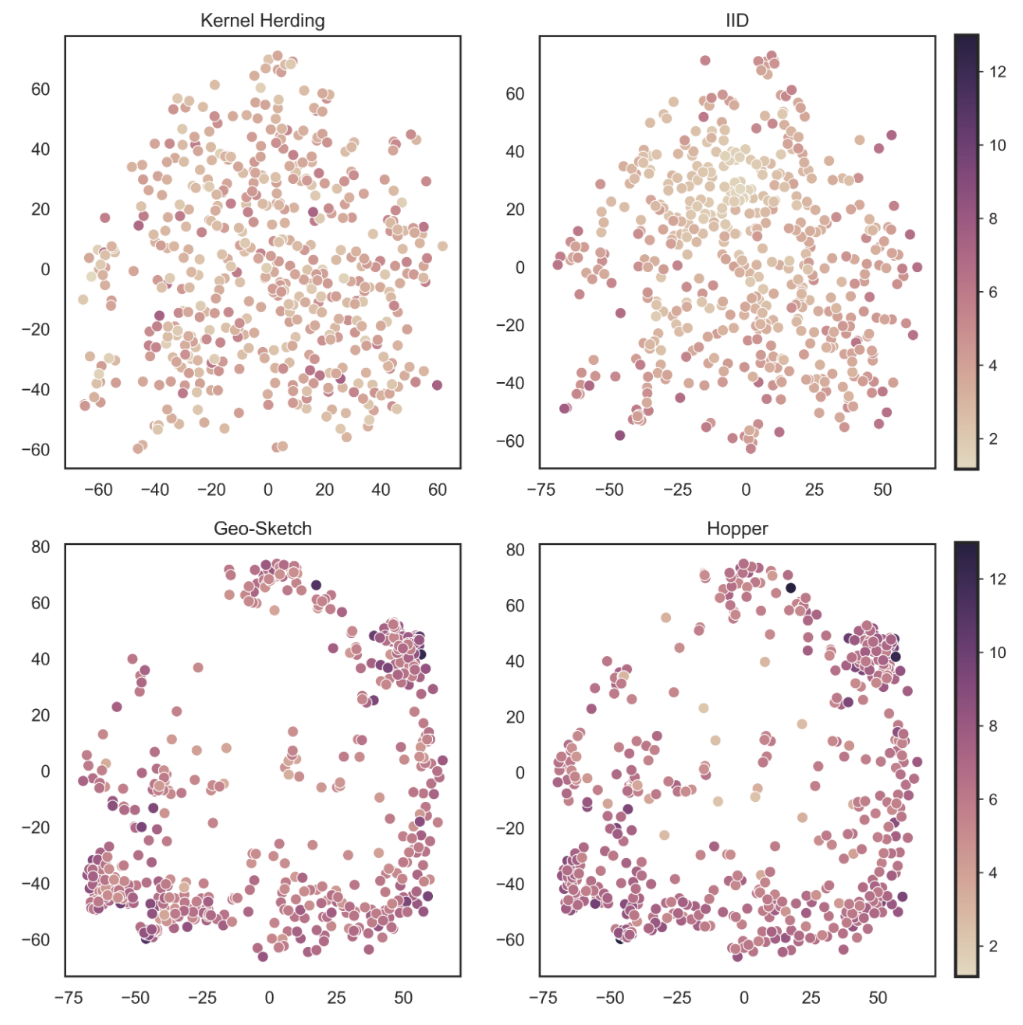}
    \caption{\bf t-SNE visualizations of cells sketched from the NK-Cell dataset colored by their the third nearest neighbor distances in the original protein-marker feature space. %Overall, distances from Geo-Sketch and Hopper sketches are relatively higher in dense regions indicating that these regions of the cellular landscape are over-represented.
    }
    \label{fig:nnbr}
\end{figure}

We provide additional context to the aforementioned t-SNE projections (Fig.~\ref{fig:cluster_tsne}) with a volumetric analysis. In particular, as t-SNE projections may not be volumetrically preserving (areas in the $2d$ space need not scale to areas in the original space), it may be the case that sketched points seem overly warped and are not representative of their coverage in the original high-dimensional protein marker feature space. To obtain a clearer view of the representation of sketched cells in the original space, we visualize the third nearest neighbor distance in the original space when scattering the cells in the $2d$ t-SNE space (Fig. \ref{fig:nnbr} plots this for a sample-set in the NK-Cell dataset). That is, dark-colored points stem from \emph{sparsely} populated regions in the original space, since the nearest neighbor of such points in the original space was \emph{far}. Similarly, light-colored points stem from \emph{densely} populated regions in the original space. As suspected above, Geo-Sketch and Hopper sketches are concentrated in \emph{sparse} regions of the space. While this acceptable for certain applications, the resulting sketches largely ignore denser regions of the cell-space, which prevents sketches from acting as a \emph{stand-in} for downstream analysis. In contrast, we observe that Kernel Herding obtains good coverage of the original space, while still including cells from sparser regions.

% The nuanced differences between all sketching methods were further explored for the NK-Cell dataset in Fig. \ref{fig:nnbr}, by quantifying the distances of each sketched point to their third nearest neighbor in the high-dimensional protein marker feature space. Moreover, cells are colored by this distance (with a darker color implying greater distance). Collections of dark-colored points that are well-dispersed in the original sample-set, but are close to each other in the sketch indicates the over-representation of a generally sparse region of the cellular landscape. This is explicitly illustrated in the tSNE plots corresponding to the Geometric Sketching and Hopper sketches (bottom row of Fig. \ref{fig:nnbr}) where the shade of samples is much darker in many regions. While sketches through IID sampling also perform well at covering dense regions, it isn't guaranteed to sufficiently represent rare populations due to its inherent stochastic nature. In contrast, sketching with Kernel Herding strikes a balance between representing rare cell-populations and preserving their frequencies. 

% \begin{figure}[H]
%     \centering
%     \includegraphics[width=1\columnwidth,scale=0.75]{figures/ToUse/all_3rd_nnbr_histogram.png}
%     \caption{\bf Histograms of the 3rd nearest neighbor distances visualized in Figure \ref{fig:nnbr} t-SNE above.}
%     \label{fig:nnbr_hist}
% \end{figure}

\subsection{Random Function Fidelity}

\begin{figure}[H]
    \centering
    \includegraphics[width=0.9\columnwidth]{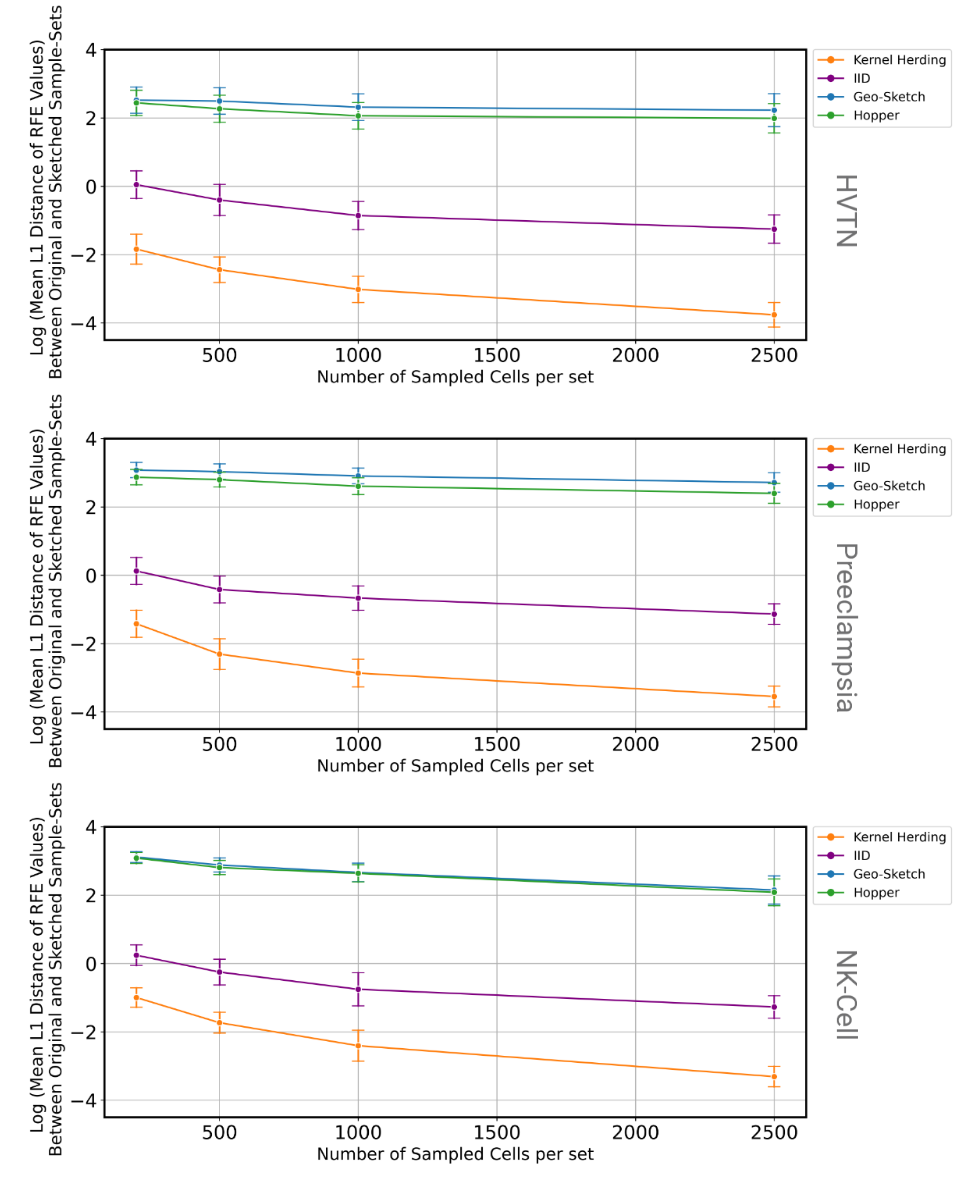}
    \caption{{\bf Random Function Evaluation} We evaluated a random function using the full and sketched versions for the HVTN, Preeclampsia and NK-Cell datasets, for sketch sizes between 200 and 2500 cells. Sketches with Kernel Herding generally produce random function evaluations (RFEs) that are the most similar to that obtained when using all cells.}
    \label{fig:rand}
\end{figure}

As previously discussed, our aim is to produce sketches that may act as a stand-in for general analysis of samples. To test how well sketches estimate the output of a \emph{wide-range} of analyses, we begin by evaluating randomly generated functions on sketches and comparing their outputs to that of the original sample (Fig.~\ref{fig:rand}).

%\noindent {\bf I. Preserving Structure of the Cellular Landscape} \\
%-two types of figures here: \\
%- Show preservation of singular values \\
%- Show preservation of cell-population frequencies

We generate random cell-wise functions, $f:\mathbb{R}^d \mapsto \mathbb{R}$, that are evaluated on sets, $\mathcal{X}$, as : $\mathbf{f}(\mathcal{X}) = \frac{1}{|\mathcal{X}|} \sum_{x \in \mathcal{X}} f(x)$. 
We compare the function evaluation on the original sample, $\mathcal{X}$, to a sketched sub-sample, $\hat{\mathcal{X}}$, using the $\ell_1$ distance: $|\mathbf{f}(\mathcal{X})-\mathbf{f}(\hat{\mathcal{X}})|$. We parameterize functions through random features \eqref{eq:randfeat}: $f_{W,\beta}(x) = \varphi_w(x)^T \beta$, where $\varphi_w(x)$ are the random features w.r.t.~a drawn set of frequencies, and $\beta$ are coefficients. Note that $f_{W,\beta}(x)$ is a \emph{highly non-linear} function in $x$. Thus, testing the discrepancies  $|\mathbf{f}_{W,\beta}(\mathcal{X})-\mathbf{f}_{W,\beta}(\hat{\mathcal{X}})|$ between multiple $W,\beta$ shall give a robust measure of the representative power of sketches.
For each dataset, we draw 5 random functions (by drawing $\beta, W$ at random) and report the average $\ell_1$ discrepancy for sketches produced with each respective method for various sketch cardinalities (see Fig.~\ref{fig:rand}, note the log-scale). Perhaps not surprisingly in light of our qualitative analysis, we see that Geo-Sketch and Hopper yield sketches that are poor stand-ins for this task. Interestingly, we also see a large advantage in Kernel Herding sketches to IID sketches. This highlights that although IID sketches retain distributional properties, Kernel Herding sketches provide a more efficient and accurate synthesis.

\subsection{Singular Value Fidelity}

Above we studied sketches' ability to act a stand-in for general non-linear evaluations on original sample-sets. Here, we now consider a specific analysis based on singular values. In particular, we study how well the singular values of the original $n \times d$ (cell $\times$ protein marker matrices) sample compares to the singular values of corresponding $m \times d$ sketches with the $\ell_1$ metric: $||\frac{1}{\sqrt{n}} \vec{\sigma}(\mathcal{X}) - \frac{1}{\sqrt{m}} \vec{\sigma}(\hat{\mathcal{X}})||_1$, where $\vec{\sigma}$ is the vector of corresponding singular values to sample or sketch. Note that this $\ell_1$ metric directly relates to differences of eigenvalues of corresponding covariance matrices, and hence is indicative of how well sketches may act as a stand-in for linear subspace analyses such as PCA.

\begin{figure}[]
    \centering
    \includegraphics[width=0.85\columnwidth]{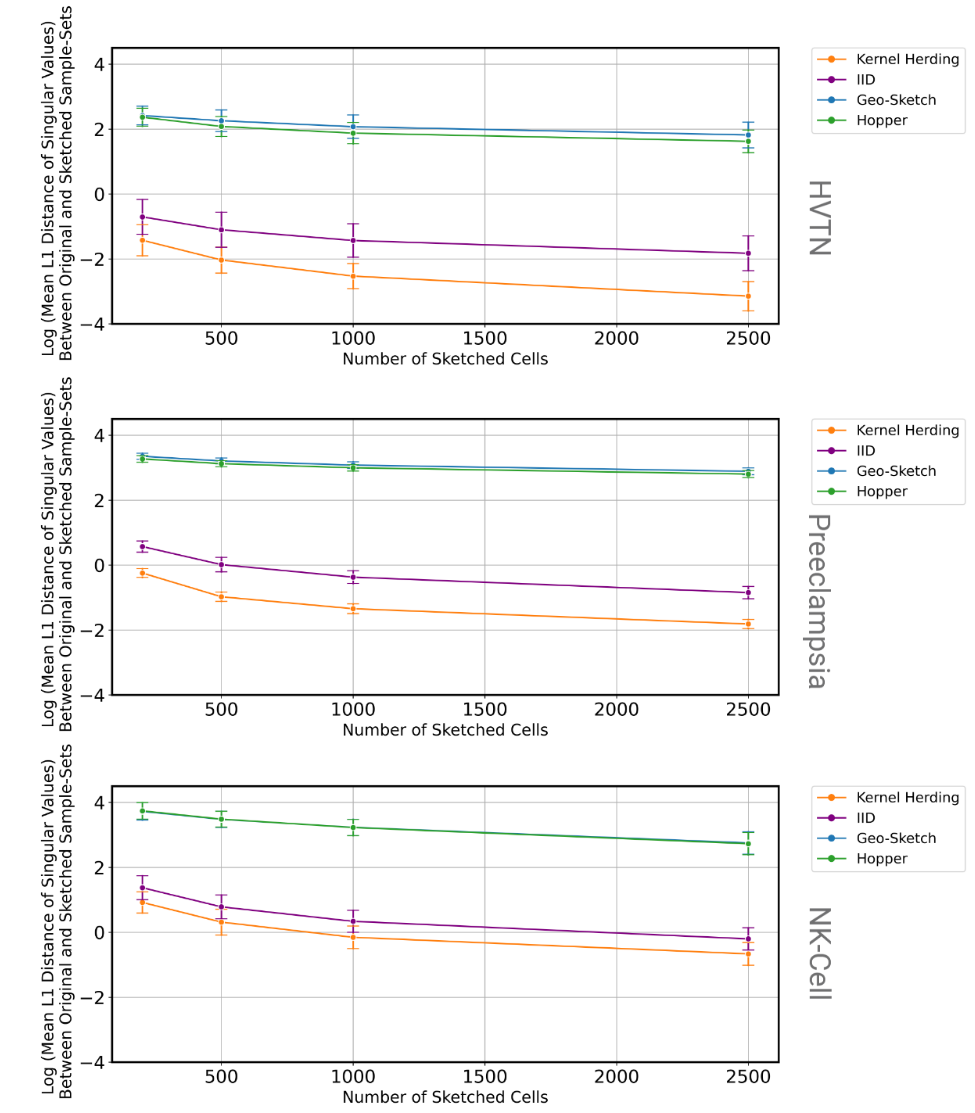}
    \caption{{\bf Singular Value Distribution Evaluation} Differences in singular value distributions of cell $\times$ protein marker matrices as quantified with an L1-norm were used to summarize the quality of sketches over a range of sketch sizes, for preserving the overall cellular landscape in the HVTN, Preeclampsia and NK-Cell datasets (top, middle, bottom, respectively). 
    %Singular values of the cell $\times$ protein marker matrices obtained through Kernel Herding are most similar to the original, non-sketched data.
    }
    \label{fig:SV}
\end{figure}

As before, we see that Geo-sketch and Hopper sketches are also unable to act as stand-ins for a singular-value analysis; a problem that does not improve as sketch sizes increase. Given that many single cell analyses rely on capturing the lower dimensional structure in samples (e.g. \cite{mefisto,phate}), this underlying bias in previous sketching approaches is limiting to their ability to act as reliable stand-ins. Moreover, we similarly observe that our proposed Kernel Herding sketches act as more faithful stand-ins to the naive IID sketches.
%To quantify the overall similarities between the original and sketched cellular landscapes, we compared the singular values of the cell $\times$ protein marker matrices for the original and sketched sample-sets. We reasoned that a quality sketch should contain a similar signal as the original sample-sets. In Figure \ref{fig:SV}, we show the similarity in singular value distributions between the original and sketched sample-sets in the HVTN, Preeclampsia, and NK-Cell datasets, respectively. Our results show that Kernel Herding produces sketches with singular value distributions that are the most similar to the original sample-sets, in comparison to the other sketching approaches. 

\subsection{Cluster Frequency Fidelity}
\label{sec:freq}

\begin{figure}[b!]
    \centering
    \includegraphics[width=0.98\columnwidth]{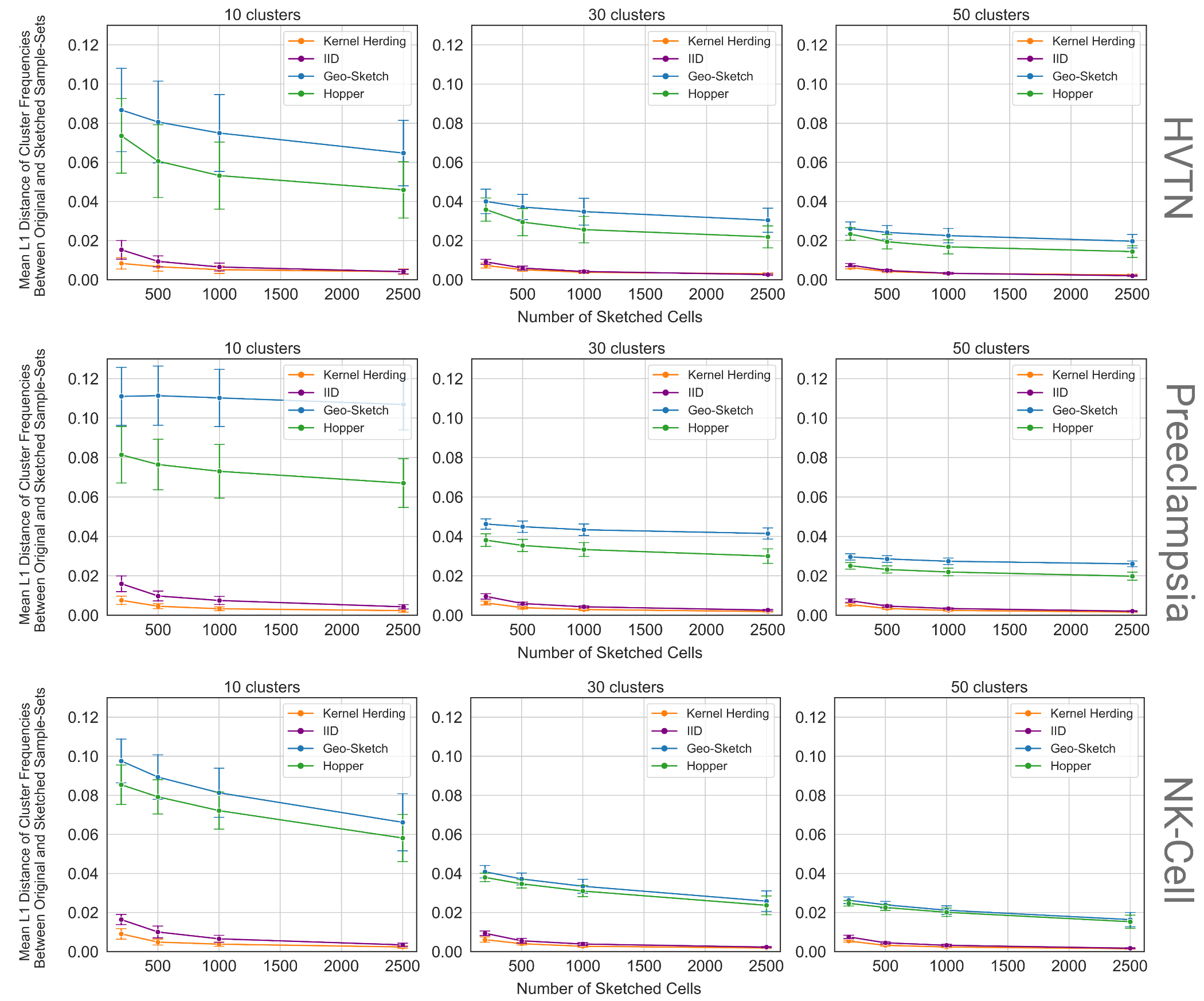}
    \caption{{\bf Cell-Population Frequency Evaluation} $k$-means was used to partition sketched sample-sets (between 200 to 2500 cells) into cell-populations (10, 30, 50 clusters). In comparison to IID, Geometric Sketching, and Hopper, Kernel Herding sketches most closely preserve the frequencies of cell-populations observed using all cells in the sample-set.}
    \label{fig:freq}
\end{figure}

Next, we study sketches' abilities to retain the overall cellular frequencies that were found in the original sample. Cell-type frequencies can be crucial for clinical outcome prediction and biological interpretation \cite{citrus,vopo}; thus, a sketch's ability to retain cell-type frequencies is pivotal for its ability to act as a stand-in to the original set in many impactful tasks.
Here we automatically detect cell-populations through an unsupervised, kmeans cluster analysis. All cells in the original sample-set are used to compute the cluster centroids, which then provide the cluster association for cells from sketches of that set. Once the centroids are computed for the original sample, each sketch is summarized according to the frequencies of the clusters in that sketch. That is, for each sketch, we compute the portion of its cells that were assigned to each cluster, $\rho(\hat{\mathcal{X}}) \in [0,1]^K$, $\rho_k(\hat{\mathcal{X}}) =  \frac{1}{|\hat{\mathcal{X}}|} \sum_{x \in \hat{\mathcal{X}}} \mathbb{I}\{k = \mathrm{argmin}_c ||x-\nu_c||\} $, where $\nu_1, \ldots, \nu_K$ are the cluster centroids. Similarly to above, we may compare the outcomes of analyzing the original sample-set, $\mathcal{X}$, to that of a corresponding sketch, $\hat{\mathcal{X}}$, with an $\ell_1$ metric: $||\rho(\mathcal{X})-\rho(\hat{\mathcal{X}})||_1$ and average this distance over all sets in the dataset.

We plot the $\ell_1$ discrepancies in frequencies for various sketching cardinalities and number of clusters in Fig.~\ref{fig:freq}. Following the same major pattern to previous experiments, we see that alternative sketching approaches (Geo-Sketch and Hopper) were unable to properly act as a stand-in to computing the cell-population frequencies in the original set. This trend is also not assuaged by increasing those sketch's sub-sampling size. Kernel Herding avoids these issues, whilst also providing a better representative sketch than IID sub-sampling. We see similar results across various number of clusters, which studies the sketches' fidelity with different granularities of cell-populations.

\begin{figure*}[t!]
    \centering
    \includegraphics[width=1.75\columnwidth]{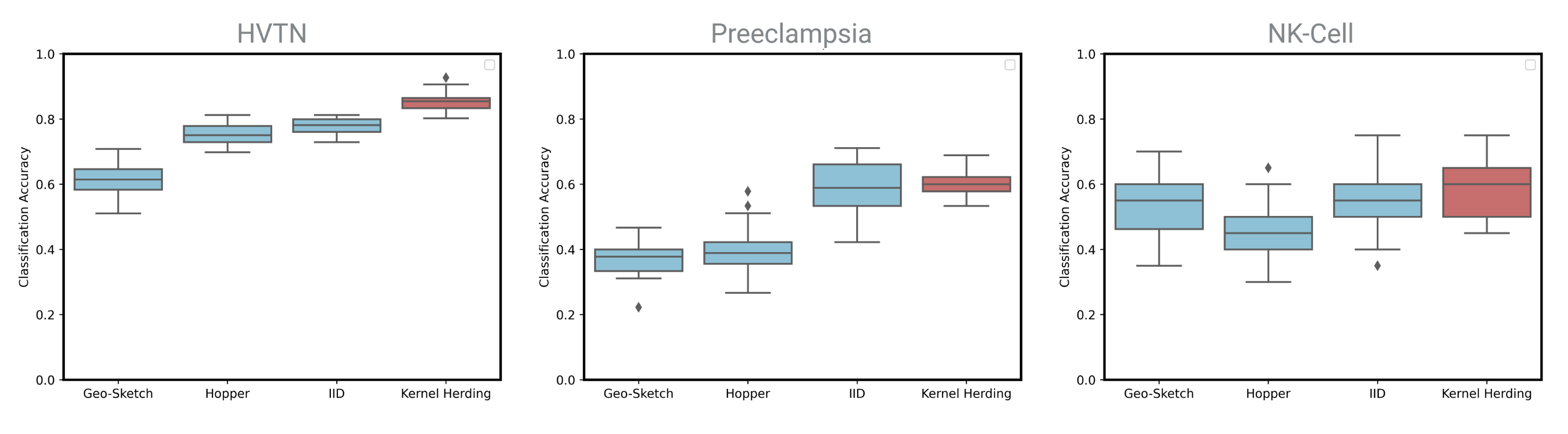}
    \caption{\emph{Clinical Outcome Classification Accuracy.}\ Sketches of 500 samples per sample set were obtained each with Geometric Sketching, Hopper, IID subsampling, and Kernel Herding and cells were partitioned amongst 30 clusters forming cell-populations. The cell-population frequencies were used as features to predict the clinical outcomes of the associated individuals for each sample-set in the HVTN, Preeclampsia, and NK-Cell datasets (left, middle, and right, respectively). We performed leave-one-out cross validation experiments for this classification task. Kernel Herding and IID produce sketches and associated features that are more predictive than those obtained through Geometric Sketching. Kernel Herding also significantly reduces the variance in classification accuracy in the Preeclampsia dataset.}
   \label{fig:class}
\end{figure*}

\subsection{Classification Effectiveness}

Finally, we explicitly test to see if the cell-population discrepancies found in Fig.~\ref{fig:freq} are relevant to producing immunological features that are useful for classifying samples according to the clinical outcomes of their associated individuals. That is, here we explicitly test that clinically relevant cell populations (such as \emph{rare-cells}) are maintained in sketches. We hypothesized that the sketches with more accurately represented frequencies (e.g. as in Sec.~\ref{sec:freq}) may be more predictive of a clinical outcome of interest. 

%Finally, we tested the ability for sketches to produce immunological features that are useful for classifying sample-sets according to the clinical outcomes of their associated individuals. Similar to our experiments that aim to evaluate the extent to which cell-population frequencies are preserved in a sketch, we hypothesized that the sketches with more accurately represented frequencies may be more predictive of a clinical outcome of interest. 

As in Sec.~\ref{sec:freq}, for each of the three cytometry datasets, cells were clustered into one of thirty clusters. Here the clusters are found from an aggregate collection of the multiple sketched sample-sets in each dataset. That is, the clusters are determined via a concatenation of all the sketched cells from all the sample sets that are in respective training sets. The frequency, or proportion of cells assigned to each population, $\rho(\hat{\mathcal{X}})$, was used as input for a downstream classification task to predict the clinical outcome for each sketched set. Note that here we trained a classifier based on multi-sample dataset $\left\{(\rho(\hat{\mathcal{X}_i}), y_i)\right\}_{i=1}^N$, where  $\hat{\mathcal{X}_i}$ is the sketch for the $i$-th sample-set, and $y_i$ is the corresponding label. Classification experiments were performed in the HVTN, Preeclampsia and NK-Cell datasets (left, middle, and right of Figure \ref{fig:class}, respectively) by splitting individuals according to leave-one-out cross validation and training an RBF support vector machine (SVM). 
Surprisingly, we found that notwithstanding the explicit focus on maintaining rare-cell types, Geo-sketch and Hopper were unable to produce sketches that lead to better accuracies than IID sub-sampling in this setting. In contrast, we found that our proposed Kernel Herding sketches lead to higher mean accuracies than IID sketching.
This held true for other cluster sizes as well (K=15,K=50). Our results show that in general, sketching with Kernel Herding produces sketches that are more clinically predictive than the baseline methods. 
This is especially apparent in the %HVTN dataset and in the 
preeclampsia dataset, where IID sketching produces highly variable classification accuracy.

\subsection{Single-cell RNA Sequencing Fidelity}
\begin{figure*}
    \centering
    \includegraphics[width=1.75\columnwidth]{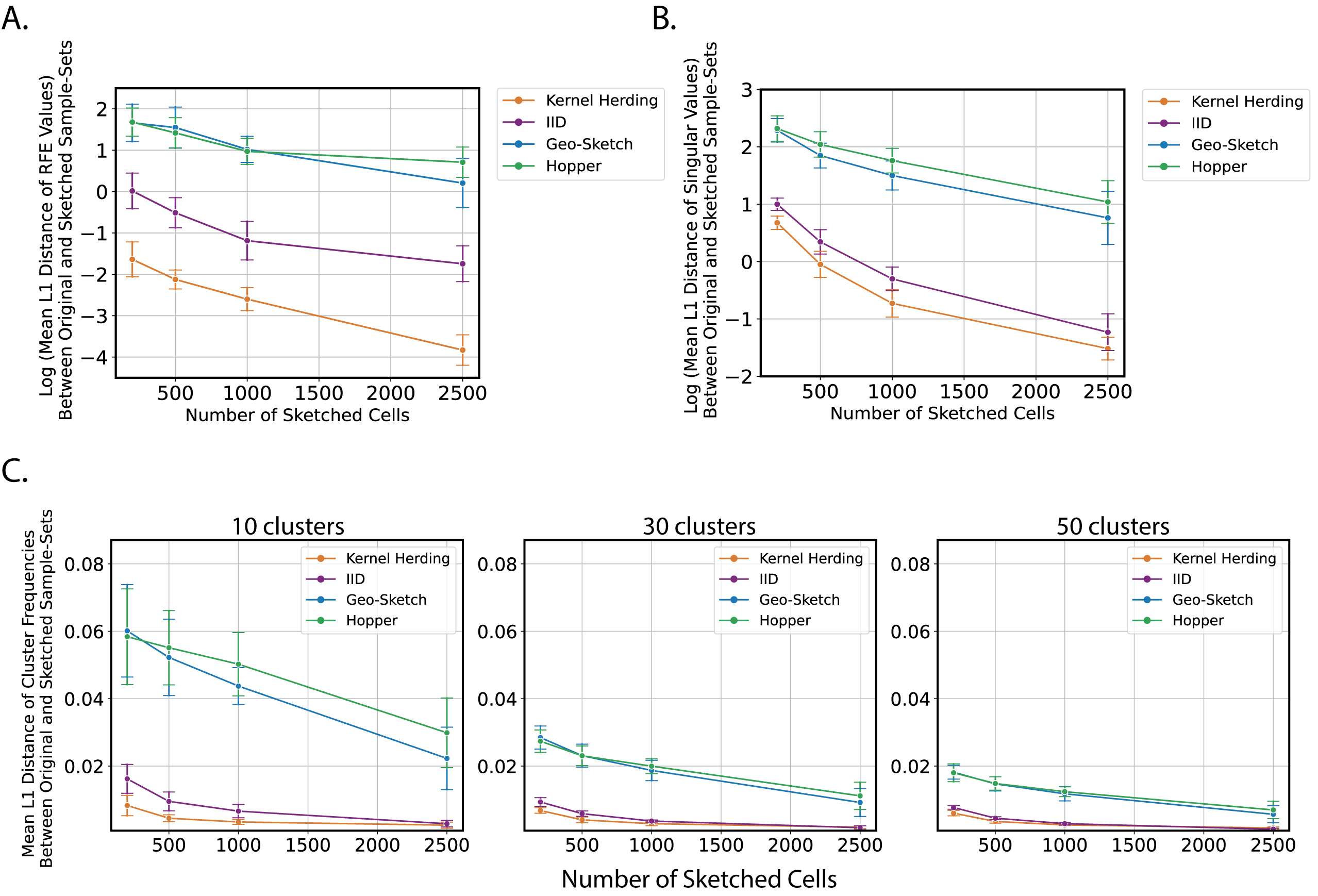}
    \caption{{\bf Single-cell RNA Sequencing Evaluation} We evaluated Hopper, Geometric Sketching, IID, and Kernel Herding on producing sketches that preserve the overall cellular landscape of single-cell RNA sequencing data. Method performance was quantified using an L1 norm between random function evaluations of original and sketched samples (A), an L1 norm between singular value distributions (B), and an L1 norm between true and sketched cell-population frequencies (C) across a range of sketch sizes (200 - 2500). Sketches with Kernel Herding produce more similar random function evaluations, similar singular value distributions, and best preserve the cell population frequencies that are observed when using all cells.}
   \label{fig:scRNAseq}
\end{figure*}

For a cross-modality comparison, we tested the performance of Kernel Herding to the sketches obtained using Geometric Sketching, Hopper, and IID subsampling on single-cell RNA sequencing (scRNA-seq) data. This modality poses an additional challenge, as scRNA-seq data typically contains 20-30 thousand gene measurements across all cells and has a high degree of sparsity and technical noise due to capture inefficiency, amplification noise, and stochasticity \cite{Kharchenko2014-ez}. Leveraging a single cell RNA sequencing dataset of patients with multiple sclerosis (MS) and healthy controls, we tested whether sketched cells resulted in similar random function evaluations, similar singular value distributions, and gave similar cell population frequencies. As shown in Figure \ref{fig:scRNAseq}, we found that Kernel Herding produces sketches that are most closely aligned with the results obtained using the full original sample. More specifically, across a range of sketch sizes for each sample-set, Kernel Herding achieves the least $\ell_1$ distance to the random function estimates in the original dataset (Fig \ref{fig:scRNAseq}A), has the most similar singular value distributions to the original data (Figure \ref{fig:scRNAseq}B), and best preserves cell population frequencies with the least $\ell_1$ norm between true and sketched cell populations (Figure \ref{fig:scRNAseq}C). 

\begin{table}[t!]
    \centering
    \caption{Runtimes (in seconds) of sketching methods on HVTN, Preeclampsia and NK-Cell datasets (number of cells ($n$ and number of features ($d$) shown underneath in parentheses) as measured on an Intel(R) Xeon(R) Gold 6226R CPU. Note that our Kernel Herding implementation was not optimized and was coded for readability. }
    \label{tab:runtimes}
    \scriptsize
    \begin{tabular}{c|c|c|c}
    \toprule
         \textbf{Methods} &  \textbf{HVTN} & \textbf{Preeclampsia} & \textbf{NK Cell} \\
          &  \textbf{(n=200k, d=11)} & \textbf{(n=215k, d=33)} & \textbf{(n=13k, d=43)} \\

    \midrule
        IID & 0.012 $\pm$ 0.001 & 0.016 $\pm$ 0.003  & 0.004 $\pm$ 0.001 \\
        Hopper & 3.13 $\pm$ 0.12 & 4.43 $\pm$ 0.15 & 1.12 $\pm$ 0.09 \\
        Geo Sketch & 23.76 $\pm$ 0.06 & 77.92 $\pm$ 0.15 & 12.29 $\pm$ 0.06 \\
    \midrule
        Kernel Herding & 26.56 $\pm$ 4.38 & 22.59 $\pm$ 1.38 & 3.64 $\pm$ 0.35  \\
        
    \bottomrule
    \end{tabular}
\end{table}

\section{Discussion and Conclusion}
Here, we presented a distribution-based, Kernel Herding approach to select a limited number of representative cells from each sample-set.
Of particular note, we recast the focus of sketching sample-sets to providing a smaller sketch that can act as a \emph{stand-in} to the original set. That is, we explicitly and quantitatively assess sketch's ability to faithfully maintain downstream outcomes when used in place of the original set.
In contrast to existing sketching approaches, such as, Geometric Sketching \cite{berger2} and Hopper \cite{berger1}, we found that Kernel Herding strikes a powerful middle-ground between preserving rare cell-populations, while also representing all major cell-types and retaining their relative frequencies. Moreover, maintaining the overall distribution of the original sample is necessary as adequate preservation of cell-population frequencies is important for linking cellular heterogeneity to clinical phenotype or external variables of interest \cite{vopo,citrus}, and for developing novel diagnostics or prognostics \cite{ramin,pree,ganio}.
Given the modern widespread use of cytometry and scRNA-sequencing technologies in clinical applications with large patient cohorts, the presented Kernel Herding based sketching approach makes such data more manageable for downstream analysis and interpretation. We showed that Kernel Herding was effective across multiple varied single cell modalities including flow and mass cytometry, as well as scRNA-seq data.

We demonstrated the usefulness of Kernel Herding at providing sketches that can serve as a stand-in for the original sample-set by measuring the fidelity of non-linear function evaluations, singular value distributions, and cell-population frequencies between the original and sketched sample-sets.
Additionally, we showed the usefulness of sketching through Kernel Herding for downstream tasks, such as clinical outcome prediction; here, Kernel Herding had the most stable performance across datasets, number of clusters, and number of sketched cells. This consistency makes it a reliable method that can be employed to obtain a representative subset, small or large, of the original distribution of samples. The runtimes of different methods reported in Table \ref{tab:runtimes} add to the practical usefulness of Kernel Herding. 

Future work could consider the implications of sketching with Kernel Herding in a broader range of tasks required to understand single-cell datasets, such as, differential abundance analysis \cite{cydar,milo,diffcyt}, or for rapid identification of phenotype-associated cells \cite{meld}. Finally, another area of future work may focus on generating variable sketch sizes across different subsections of the cellular landscape, depending on prior knowledge or scientific question.

\begin{acks}
This research was partly funded by NSF grant IIS2133595 and by NIH 1R01AA02687901A1.
\end{acks}

\bibliographystyle{ACM-Reference-Format}
\bibliography{sample-authordraft}
\end{document}